\documentclass[%
superscriptaddress,
preprint,
 amsmath,amssymb,
aps,
]{revtex4-2}

\usepackage{graphicx}% Include figure files
\usepackage{dcolumn}% Align table columns on decimal point
\usepackage{bm}% bold math
\begin{document}

%\preprint{APS/123-QED}

\title{Externally-triggerable optical pump-probe scanning tunneling microscopy with a time resolution of tens-picosecond}

\author{Katsuya Iwaya}
 \email{katsuya\_iwaya@unisoku.co.jp}
 \affiliation{UNISOKU Co., Ltd. Hirakata, Osaka 573-0131, Japan
 }

\author{Munenori Yokota}%
\affiliation{UNISOKU Co., Ltd. Hirakata, Osaka 573-0131, Japan
}%

\author{Hiroaki Hanada}
\affiliation{UNISOKU Co., Ltd. Hirakata, Osaka 573-0131, Japan
}%

\author{Hiroyuki Mogi}
\affiliation{ 
Faculty of Pure and Applied Sciences, University of Tsukuba,
Ibaraki 305‑8573, Japan
}% 

\author{Shoji Yoshida}
\affiliation{ 
Faculty of Pure and Applied Sciences, University of Tsukuba,
Ibaraki 305‑8573, Japan
}% 

\author{Osamu Takeuchi}
\affiliation{ 
Faculty of Pure and Applied Sciences, University of Tsukuba,
Ibaraki 305‑8573, Japan
}% 

\author{Yutaka Miyatake}
\affiliation{UNISOKU Co., Ltd. Hirakata, Osaka 573-0131, Japan 
}% 

\author{Hidemi Shigekawa}
\email{hidemi@ims.tsukuba.ac.jp}
\affiliation{ 
Faculty of Pure and Applied Sciences, University of Tsukuba,
Ibaraki 305‑8573, Japan
}% 

\date{\today}% It is always \today, today,

\begin{abstract}
Photoinduced carrier dynamics of nanostructures play a crucial role in developing novel functionalities in advanced materials. Optical pump-probe scanning tunneling microscopy (OPP-STM) represents distinctive capabilities of real-space imaging of such carrier dynamics with nanoscale spatial resolution. However, combining the advanced technology of ultrafast pulsed lasers with STM for stable time-resolved measurements has remained challenging. The recent OPP-STM system, whose laser-pulse timing is electrically controlled by external triggers, has significantly simplified this combination but limited its application due to nanosecond temporal resolution. Here we report an externally-triggerable OPP-STM system with a temporal resolution in the tens-picosecond range. We also realize the stable laser illumination of the tip-sample junction by placing a position-movable aspheric lens driven by piezo actuators directly on the STM stage and by employing an optical beam stabilization system. We demonstrate the OPP-STM measurements on GaAs(110) surfaces, observing carrier dynamics with a decay time of $\sim170$~ps and revealing local carrier dynamics at features including a step edge and a nanoscale defect. 
The stable OPP-STM measurements with the tens-picosecond resolution by the electrical control of laser pulses highlight the potential capabilities of this system for investigating nanoscale carrier dynamics of a wide range of functional materials.
\end{abstract}

\maketitle

\section{Introduction}

The ability to measure carrier dynamics in nanoscale materials and devices is an important capability that requires experimental techniques with both high spatial and high temporal resolutions\cite{Shah}. To this end, many time-resolved techniques in combination with methods such as electron microscopy\cite{Zewail,Feist,Ideta}, photoemission electron microscopy\cite{Fukumoto,Man}, and x-ray diffraction\cite{Eichberger} have been reported.
Scanning tunneling microscopy/spectroscopy (STM/STS) is a powerful technique to probe topographic and spectroscopic properties of various material surfaces with high spatial and energy resolutions. However, the temporal resolution of conventional STM is limited to the sub-millisecond range by the preamplifier bandwidth ($\sim 1 $~kHz). To overcome this limitation, considerable effort has been made since its invention\cite{Mamin,Weiss,Wintterlin,Kemiktarak}. Among these, the application of optical pump-probe (OPP) techniques to STM can circumvent the limitations of the circuit bandwidth, achieving higher temporal resolutions\cite{Hamers,Nunes,Weiss_APL,Keil,Grafstrom}. 

An OPP-induced tunneling current is generally weak to detect so that we need to employ a modulation technique using a lock-in amplifier. However, the modulation of optical intensity causes severe problems such as thermal expansions of STM tip and sample. Since changes in the tip-sample distance are exponentially multiplied in the tunneling current, such conventional OPP methods cannot be directly combined with STM.
In 2004, an exquisite delay-time modulation technique 
to suppress the thermal expansion effect has been invented\cite{Takeuchi_APL}. With subsequent noise-level and delay-time improvements\cite{Terada_NatPhoton,Patent1}, the OPP-STM is now capable of probing the nonequilibrium dynamics of systems such as the atomic scale carrier dynamics around a single impurity on GaAs(110) surface\cite{Yoshida_APEX,Kloth_SciAdv}, the visualization of the ultrafast carrier dynamics in a GaAs-PIN junction\cite{Yoshida_Nanoscale}, and the relaxation dynamics of polarons bound to oxygen vacancies on rutile TiO$_2$(110) surface\cite{Guo}. 
Furthermore, recent studies have realized another time-resolved STM utilizing a sub-cycle electric field as bias voltage between the STM tip and sample, called electric-field-driven STM.
By measuring an instantaneous tunneling current induced by the sub-cycle electric field, ultrafast time-resolved measurements can be performed. 
The electric-field-driven STM enables the temporal resolution faster than 1~ps and 30~fs while maintaining the spatial resolution of STM using terahertz (THz) and mid-infrared pulses\cite{Cocker_NatPhoton,Cocker_Nature,Li,Garg,Yoshida_ACSPhoton,Wang, Arashida_ACSPhoton,Arashida_APEX}. 
These efforts have substantially expanded the possibilities of time-resolved STM. However, the use of sub-cycle pulsed 
electric fields still requires various expertise, including the creation and control of electric fields.

The recent OPP-STM systems, whose laser-pulse timing is electrically controlled by external triggers, have significantly improved both the ease of use and the stability of the optical system\cite{Guo,Mogi_FPGA,Kloth_RSI}, 
but the temporal resolutions have been limited to the nanosecond range.
For example, recent time-resolved observation of exciton dynamics in transition metal dichalcogenides has revealed the lifetime of exciton in several tens picosecond ranges\cite{Mogi_npj}.  It is thus indispensable to improve the temporal resolution higher than the nanosecond range. 

In addition, in most previous OPP-STM systems, the focus lens is placed at a viewport on the ultrahigh vacuum (UHV) chamber while hanging the STM stage by springs for vibration isolation.
This configuration causes unstable laser illumination of the tip-sample junction because relative positions of the sample and the lens are affected by vibration noises.
The change in the light intensity due to the unstable laser illumination causes unexpected problems such as the thermal expansion effect, making it difficult to observe physical phenomena of interest correctly. 
Therefore, to utilize this experimental technique in a wider range of research fields, further improvements in the stability of the laser spot with a high temporal resolution are strongly required.

In this study, we report the design and performance of a newly developed externally-triggerable OPP-STM system that enables us to conduct tens-picosecond time-resolved measurements with long-term stability. 
To simplify the optical system, we employ externally-triggerable picosecond laser systems (pulse width $\sim45$~ps) and control the timing of laser pulses electrically. 
We also show that by placing the aspheric lens on the STM stage and using an optical beam stabilization system, the position of the laser spot on the sample surface is stable for hours, thus suitable for long-term experiments.
The OPP-STM system with both high temporal resolution and high optical stability will facilitate the widespread use of this method for understanding nanoscale carrier dynamics.

\section{Experimental methods}
\subsection{System overview}

The newly developed OPP-STM system used in this study consists of three components: the low temperature UHV STM system, the optical system that includes two lasers and the beam stabilization system, and the delay-time control system  [Fig. 1].

The STM system is based on the UNISOKU USM1400 model and is composed of three UHV chambers: a load-lock chamber for tip and sample exchange, a tip and sample preparation chamber, and a STM  observation chamber [Fig.~2(a)]. The three chambers are mounted on a passive vibration isolation table (ADF-1311YS, Meiritz Seiki Co., Ltd.). The preparation and observation chambers are pumped by two ion pumps (240 and 125 L/s), each equipped with a titanium sublimation pump. The base pressure of the observation chamber is $8\times10^{-8}$~Pa at room temperature and $2\times10^{-8}$~Pa when the STM head is cooled down. The base temperature at the STM stage when cooled with liquid nitrogen and liquid helium is 78~K and 6~K, respectively.

\begin{figure*}
\includegraphics[width=15cm]{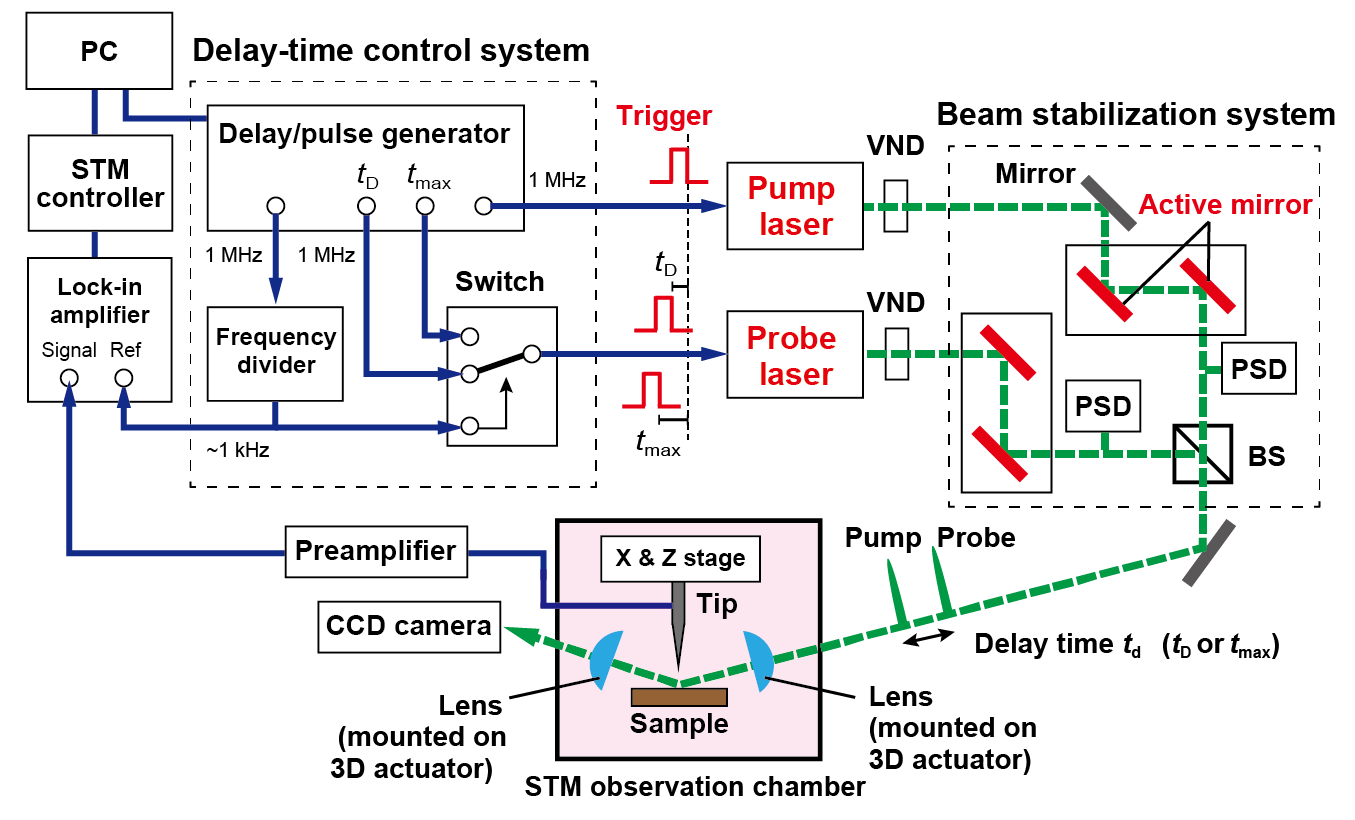}% Here is how to import EPS art
\caption{Schematic diagram of the OPP-STM system. PSD: position sensitive detector, BS: beam splitter, VND: variable neutral density filter.}
\label{fig1}
\end{figure*}

Both the tip and the sample are aligned horizontally in the STM head [Fig.~3(a)], and are transferred to the STM head from the top using a vertical transfer rod. The flag-type sample holder is loaded onto the sample stage which is fixed without moving capability. The tip coarse approach and the tip lateral position control are driven by the stick and slip motion of shear piezo stacks (six stacks with travel distance of $\pm 2.5$~mm for coarse approach, and three stacks with travel distance of $\pm 3$~mm for lateral positioning). The STM stage is vibrationally isolated by an eddy current damper that is composed of four stainless-steel springs and nine samarium–cobalt magnets.
The absence of significant noise peaks up to 1~kHz is confirmed in a tip-sample distance noise spectrum (Supplementary Fig.~S1 online).  

The aspheric lens focusing the laser light on the sample surface is located close to the sample stage [“Lens \#1” in Fig.~3(a), A397-A, Thorlabs, Inc.], and the light enters with the incident angle of 55$^\circ$ normal to the sample surface. With this configuration, the numerical aperture of 0.3 is obtained (the working distance is 9.64~mm). To precisely adjust the laser spot on the sample surface, the aspheric lens is mounted on a home-made 3D piezoelectric actuator that moves in the x, y, and z directions. The movement in each direction is driven by the stick and slip motion of three shear piezo stacks. The travel distance in the x, y, and z directions are $\pm 3$~mm, $\pm 3$~mm, and $\pm 2$~mm, respectively. Another aspheric lens, similarly mounted on the 3D piezoelectric actuator [“Lens \#2” in Fig.~3(a)], is used to observe the tip apex, the sample surface, and the focused laser spot {\it in situ} through a CCD camera [Fig.~3(b)].  
The optical magnification by the objective lens ("Lens \#2") and imaging lens (located outside the UHV chamber, AC254-200-A, Thorlabs, Inc.) is $\sim18$x. 

The sample surface under the STM tip is illuminated by a sequence of pump and probe pulses [Fig.~1], and the tunneling current is detected using a commercial current preamplifier ($10^9$~V/A, DLPCA-200, Femto Messtechnik GmbH). The change in the tunneling current induced by the delay-time modulation technique is measured using a lock-in amplifier as a function of the delay time $t_{\rm d}$ between the pump and probe pulses. We discuss this latter measurement in more detail later.

\begin{figure}
\includegraphics[width=8cm]{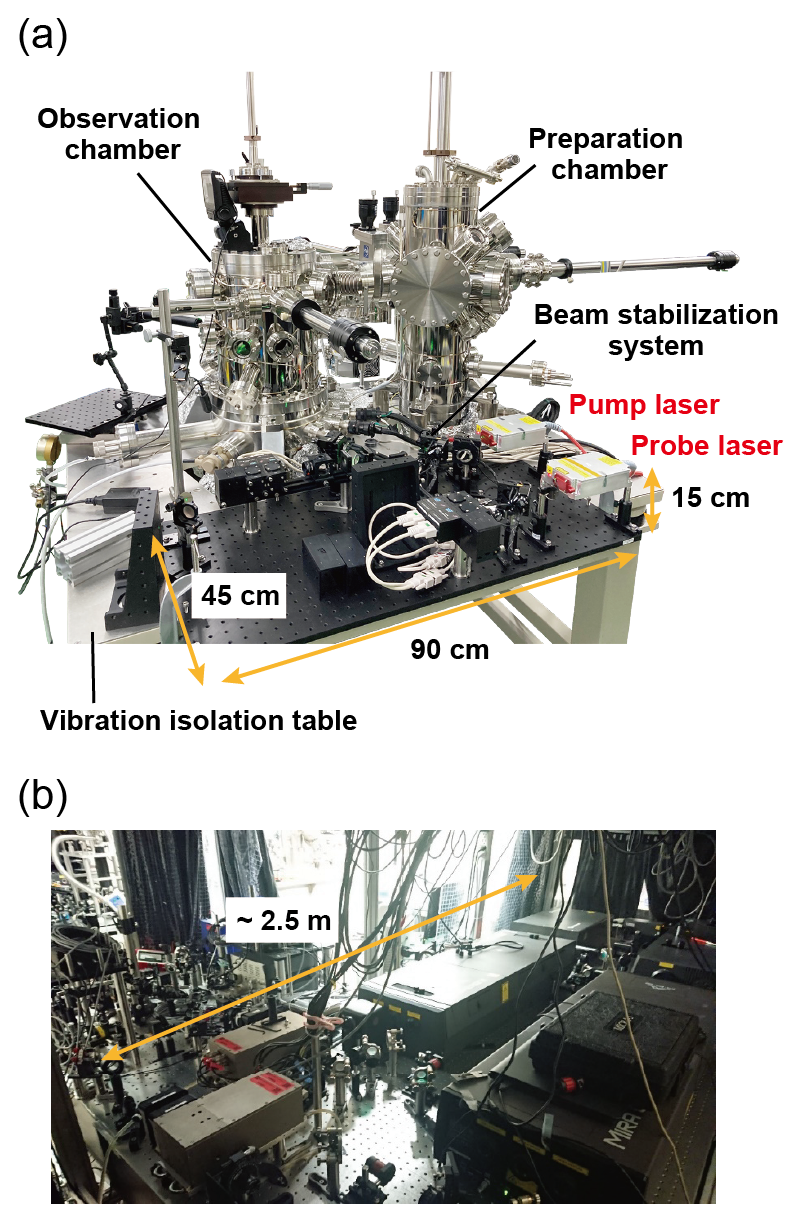}% Here is how to import EPS art
\caption{(a) Photograph of the STM and the optical system developed in this study. (b) Photograph of the conventional optical system used in ref.\cite{Terada_NatPhoton}.}
\label{fig2}
\end{figure}

\begin{figure*}
\includegraphics[width=12cm]{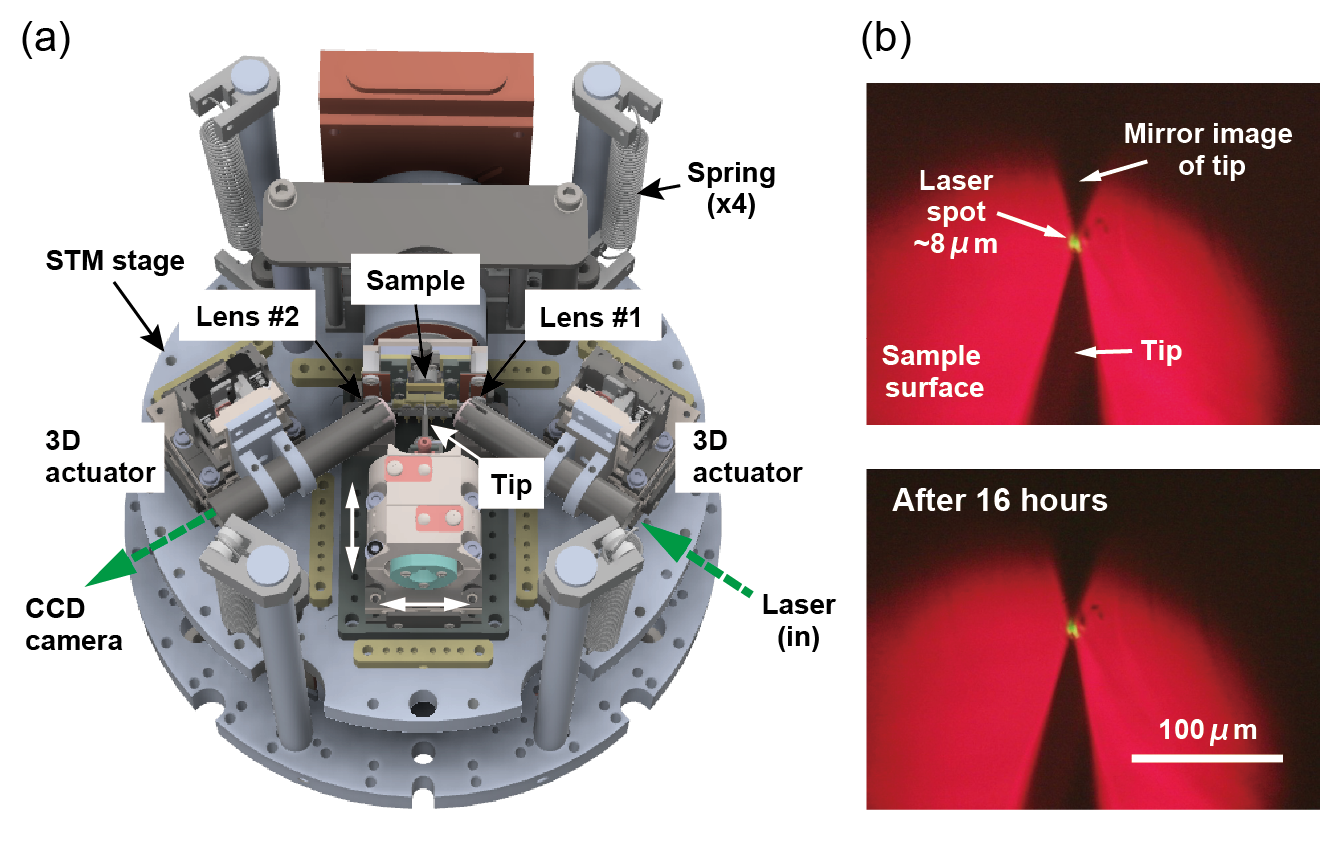}% Here is how to import EPS art
\caption{(a) Three-dimensional illustration of the STM unit (top-view). Electrical wirings and copper heat links are not shown for clarity. (b) (Top) Optical image of the tip and its mirror image on GaAs(110) surface with a laser spot illuminated at the tunneling junction. (Bottom) The image taken after 16 hours, showing the stability of the laser spot.}
\label{fig3}
\end{figure*}

\subsection{Optical system}

The optical system consists of the pump laser, the probe laser, and the beam stabilization system [Fig.~1]. To realize a system that is both compact and easy-to-use, we use two picosecond pulsed laser systems (KATANA 05, NKT Photonics), whose laser-pulse timing can be electrically controlled by external triggers. The central wavelength of the lasers is 532~nm and the pulse width is $\sim 45$~ps.

To perform stable OPP-STM measurements, the long-term stability of the laser spot on the sample surface is critical.  
To keep both the pump and the probe pulses aligned on the same axis, we use the active beam stabilization system (Aligna, TEM Messtechnik GmbH) that controls one pair of active mirrors to stabilize each of the pump laser and of the probe laser spot position on the position sensitive detector (PSD) [Fig.~1]. In addition, the two laser heads and the beam stabilization system are placed on the vibration isolation table as shown in Fig.~2(a). The total dimension of the optical system ($90~{\rm cm} \times 45~{\rm cm} \times 15~{\rm cm}$) is substantially smaller than the previous OPP system used in ref.\cite{Terada_NatPhoton} [Fig.~2(b)].
We also note that placing the aspheric lens on the same stage with the STM head is crucial to keeping the laser spot position on the sample surface unchanged for long periods of time because the relative positions of the sample and the lens are kept unaffected by vibration noises. 

To demonstrate the stability of laser spot on the sample surface, we monitor a tip and its mirror image on GaAs(110) cleaved surface together with a laser spot focused at the tip-sample junction. Figure 3(b) demonstrates that the position of the laser spot remains unchanged even after 16 hours. 
We also monitor both the tunneling current and tip height with the feedback loop closed when the laser is illuminated (laser power: 0.25~mW) for 12 hours at $T = 6$~K and confirm that the standard deviation of the tunneling current is about 0.6~\% of the average set-point current with reasonably stable tip heights (Supplementary Fig.~S1 online).
These results indicate that our optical setup is stable enough for long-term OPP-STM experiments.

\subsection{Delay-time modulation technique}
In the OPP-STM measurement, the sample surface under the tip is first excited by a pump pulse, and subsequently by a probe pulse with a delay time $t_{\rm d}$ [Fig.~1]. When $t_{\rm d}$ is sufficiently long [(1) in Fig.~4(a)], most of the photocarriers excited by the pump pulse relax to the ground state before the subsequent probe pulse is illuminated; thus, a similar number of carriers would be excited by the probe pulse as with the pump pulse, resulting in a large transient current $I^*_{\rm probe}$. 

In contrast, when $t_{\rm d}$ is substantially short, the excited states remain occupied when the probe pulse illuminates the sample so that the saturation of optical absorption occurs, resulting in a smaller $I^*_{\rm probe}$ [(3) in Fig.~4(a)]. By illuminating a pair of pump and probe pulses sequentially and by varying $t_{\rm d}$, a time-averaged tunneling current $<I>$ is detected as a function of $t_{\rm d}$ [Fig.~4(b)]. 

When $t_{\rm d} < 0$, a pump pulse and a probe pulse are swapped. A symmetric lineshape with respect to $t_{\rm d} = 0$ can be observed when the intensities of the pump and probe laser pulses are finely adjusted to be equivalent, allowing us to evaluate measurement conditions such as laser intensity and delay time settings. By combining this technique with the STM capability, the ultrafast carrier dynamics of the sample surface local structures can be investigated. 

To detect a weak OPP-induced tunneling current, we need to employ a modulation technique using a lock-in amplifier. However, the modulation of optical intensity causes severe problems such as thermal expansions of STM tip and sample as described before.
Therefore, we need the delay-time modulation technique designed to suppress the thermal expansion effect\cite{Takeuchi_APL} [Fig.~4(c)]. In this technique, we use two delay times [$t_{\rm D}$ and $t_{\rm max}$ in Figs.~1 and 4(c)]. The longer delay time $t_{\rm max}$ is generally set to a half of the laser pulse interval (for instance, 0.5~$\mu$s for 1~ MHz repetition rate), which corresponds to the longest delay time available for the selected repetition rate. We modulate the delay time between $t_{\rm D}$ and $t_{\rm max}$ at 1~ kHz and detect the resultant tunneling current $\Delta I(t_{\rm D}) = <I(t_{\rm D})> - <I(t_{\rm max})>$ using the lock-in amplifier [Fig.~4(c), bottom]. By sweeping $t_{\rm D}$ slowly along with the lock-in detection, we obtain $\Delta I$ as a function of $t_{\rm D}$, called an OPP tunneling current--delay time curve hereafter. This technique enables us to keep the thermal load at the tunnel junction constant, suppressing the thermal expansion effect substantially\cite{Kloth_RSI}.    

\begin{figure*}
\includegraphics[width=15cm]{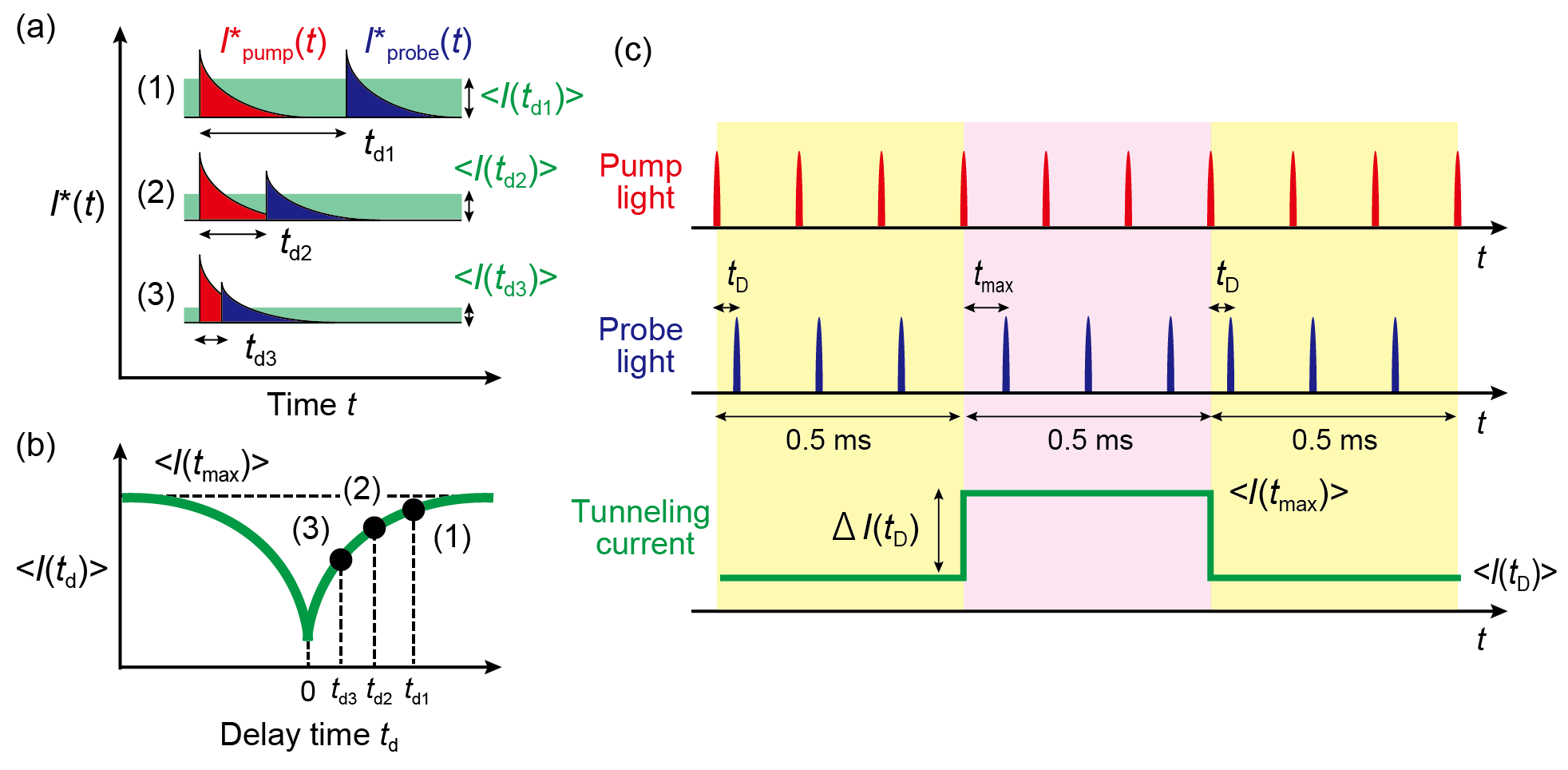}
\caption{(a) Relationship between transient tunneling current $I^*$ induced by the pump and probe light and delay time $t_{\rm d}$. Time-averaged tunneling current $<I>$ is shown for each case. (b) $<I>$ as a function of $t_{\rm d}$. The time-averaged $<I>$ corresponding to each case in (a) is plotted with corresponding number. (c) Schematic of the delay-time modulation technique. The delay time between the pump and probe pulses is modulated between $t_{\rm D}$ and $t_{\rm max}$ at $\sim 1$~kHz. Consequently, $<I>$ is also modulated between $<I(t_{\rm D})>$ and $<I(t_{\rm max})>$ at $\sim 1$~kHz, and the lock-in amplifier detects $\Delta I(t_{\rm D})= <I(t_{\rm D})>-<I(t_{\rm max})>$.}
\label{fig4}
\end{figure*}

\subsection{Delay-time control system}
To conduct the delay-time modulation technique shown in Fig.~4(c), previous studies have either used a mechanically movable mirror to change the optical path length\cite{Takeuchi_APL}, or a pulse picker such as a Pockels cell to extract specific pulse pairs\cite{Terada_NatPhoton}. These systems are inevitably large and complicated, requiring considerable skills and experience to operate. Recently, a compact tabletop OPP system, in which the timing of laser pulses is electrically controlled by external triggers using the field-programmable gate array, has been reported\cite{ Mogi_FPGA, Patent2}. However, while this approach has significantly improved the operational ease of the optical system, it still requires expertise in electronic circuits for modifications and maintenance. Thus, developing a simpler optical system both in terms of ease of operation and maintenance is important for the OPP-STM method to be widely used. 

In this study, we build the delay-time control system  by combining the following commercial products: a digital delay/pulse generator (DG645, Stanford Research Systems), a high-speed switch (HMC-C011, Analog Devices Inc.) and a frequency dividing circuit (74HC4040, Toshiba) [Fig.~1]. Here, the trigger pulse for the pump light (pump trigger) is directly input from the pulse generator to the pump laser at the repetition rate of 1~MHz. The two trigger pulses for the probe light (probe triggers) with delay time $t_{\rm D}$ and $t_{\rm max}$ relative to the pump trigger are input to the switch at the repetition rate of 1~MHz [Fig.~1]. Either the probe trigger with $t_{\rm D}$ or with $t_{\rm max}$ is selected at the frequency of $\sim 1$~kHz in the high-speed switch. Thus, a train of probe trigger pulses with $t_{\rm D}$ or $t_{\rm max}$ are alternately input to the probe laser with a time-interval of $\sim 0.5$~ms, to generate a train of probe pulses shown in Fig.~4(c). 

The 1~kHz switching signal is produced by the frequency divider, which is set to $1/1024$ $(1~{\rm MHz} \times 1/1024 = \sim 977~{\rm Hz})$. This latter frequency is also used as the reference signal for the lock-in amplifier. The time resolution of the pump-probe experiment using this system is theoretically estimated to be $\sim 70$~ps, limited by the temporal width (45~ps), the jitter of the laser pulse (15~ps), and the jitter of the electrical trigger (25~ps). 
We measure cross correlation between the pump and probe laser pulses using the sum-frequency generation method, and the correlation width estimated by Gaussian fitting is $77.6 \pm 2.1$~ps (Supplementary Fig.~S2 online) and in reasonable agreement with the theoretical value.

\section{Results}
First, we demonstrate the light-modulated scanning tunneling spectroscopy measurement \cite{Takeuchi_APL2,Yoshida_APL} on GaAs(110) surface, where we measure $I$-$V$ curves both under dark conditions (without laser illumination), and with laser illumination at a repetition rate of 1~MHz. The atomically flat and clean surface is obtained by cleaving a commercially available $n$-type (100)-oriented GaAs wafer (doped with silicon at the density of $\sim 5\times10^{17}$~cm$^{-3}$) along the (110) plane at room temperature in UHV conditions. In this study, mechanically polished PtIr tips are used for all STM measurements. A typical constant-current STM image at a sample bias voltage $V$ of $-3$~V in Fig.~5(a) shows the one-dimensional rows along the [$1\overline{1}0$], consisting of the atomic lattice of As where the atomic spacing between the rows along the [$1\overline{1}0$] direction is 0.565~nm and along the row is 0.4~nm \cite{Feenstra}. 

A typical $I$-$V$ curve taken without laser illumination is shown in blue in Fig.~5(b). A negligibly small tunneling current at $V > 0$ is attributed to a depletion layer formed at the surface due to the tip-induced band bending effect\cite{deRaad}. In the case of an $n$-type semiconductor surface, the conduction band of the sample is bent upward near the surface when $V > 0$ [Fig.~5(c), top], preventing electrons tunneling from the tip. This results in a small tunneling current at $V > 0$, as observed in Fig.~5(b). 

\begin{figure*}
\includegraphics[width=12cm]{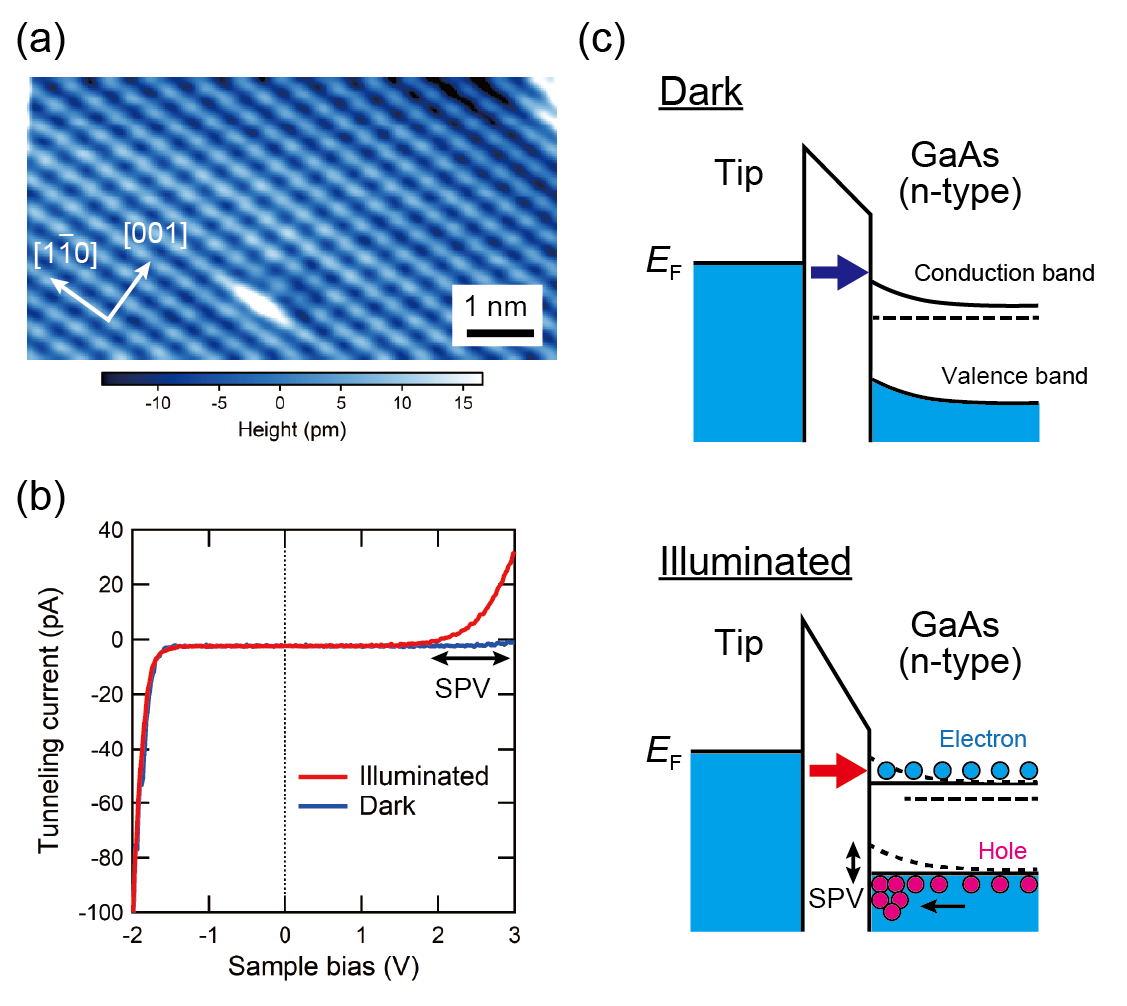}% Here is how to import EPS art
\caption{(a) Constant-current STM image of GaAs(110) surface. Set-point: sample bias voltage $V = -3$~V, and tunneling current $I_{\rm t} = 100$~pA. $T = 78$~K. (b) $I$-$V$ curves obtained with (red) and without laser illumination (blue). Set-point: $V = -2$~V, $I_{\rm t} = 100$~pA. $T = 78$~K. (c) Schematic illustrations of the band structures of the $n$-type GaAs surface when $V > 0$ without (top) and with laser illumination (bottom).}
\label{fig5}
\end{figure*}

When the sample surface is illuminated by laser pulses, electron-hole pairs are generated; according to the tip-induced surface potential, while the holes in the valence bands accumulate at the surface under the tip, the electrons in the conduction bands drift back into the bulk [Fig.~5(c), bottom]. This redistribution of photocarriers induces efficient screening of the tip potential and suppresses the upward band bending as shown at the bottom of Fig.~5(c). The energy shift due to the screening is called surface photovoltage (SPV). Consequently, the tunneling current at $V > 0$ greatly increases under illumination [red in Fig.~5(b)], and the SPV is estimated to be about 1.1~V at $V=+3$~V. This behavior is in good agreement with previous results\cite{Yoshida_APEX}, and is evidence that the sample surface under the tip is sufficiently illuminated by the laser pulses. 

Next, we demonstrate an OPP tunneling current measurement at $T = 78$~K on the GaAs(110) surface using the delay-time modulation technique described in an earlier section. Figure 6(a) shows an OPP tunneling current versus delay time curve obtained with a laser repetition rate of 1~MHz and a maximum delay time $t_{\rm max}$ of 500~ns. The laser power is 0.5~mW. The lineshape is well fitted by a combination of two exponential functions, where the decay times are estimated to be $4.5\pm0.2$ and $121.3\pm8.3$~ns. 
The temporal resolution improved in this study allows us to observe the faster decay time ($\sim 5$~ns), which is inaccessible using previous externally-triggerable optical systems\cite{Mogi_FPGA,Guo}. 
It is known that for an $n$-type semiconductor, the excited state relaxes to the original state through two processes\cite{Terada_JP}. One is the decay of the photocarriers in the bulk (bulk-side decay) via recombination, drift and diffusion [Fig.~6(b), top]. The other is the decay of the photocarriers trapped at the surface (surface-side decay) via thermionic emission [Fig.~ 6(b), bottom]. The surface-side decay time is generally longer than the bulk-side decay time because counterpart carriers are absent near the surface due to the depletion layer, and holes near the surface need to move into the bulk to combine with electrons. Thus, the two decay times obtained by fitting correspond to the bulk-side decay ($\sim 5$~ns) and the surface-side decay ($\sim 120$~ns), and are consistent with previous results\cite{Yokota}. 

\begin{figure*}
\includegraphics[width=12cm]{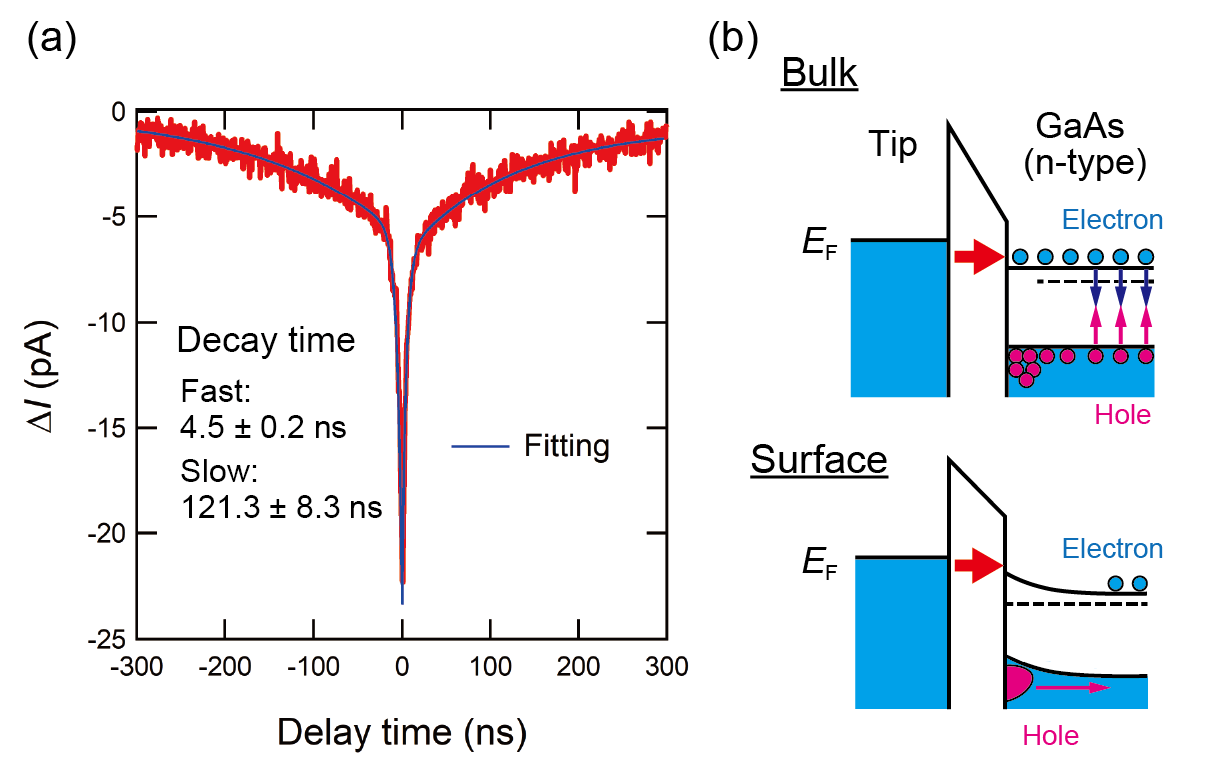}% Here is how to import EPS art
\caption{(a) Typical OPP tunneling current--delay time curve of GaAs(110) surface. Set-point: $V = +2.8$~V, $I_{\rm t} = 100$~pA. Laser power is 0.5~mW. $T = 78$~K. The step of delay time is 795~ps, and the averaging time at each delay time is 30~ms. The curve is averaged over 10 sweeps and fitted with two exponential functions.
(b) Schematic illustration of the band structures of the $n$-type GaAs during the time-resolved measurement when $V > 0$. The decay processes of the photocarriers in the bulk (top) and the surface (bottom) are shown.
}
\label{fig6}
\end{figure*}

Next, we demonstrate a grid-point OPP tunneling current measurement at $T = 6$~K. Figure 7(a) shows a constant-current $50~{\rm nm} \times 50~{\rm nm}$ STM image of a GaAs(110) surface where a nanoscale protrusion (possibly produced after cleaving, referred to as “bump structure” hereafter) and a step edge are observed over flat terraces. OPP tunneling current--delay time curves are measured at $50 \times 50$ grid points in the same field of view, each point taking $\sim 30$ seconds for a total of $\sim 21$ hours. Representative OPP tunneling current--delay time curves taken at the bump structure, the step edge and the terrace are shown in Fig.~7(b). For comparison, a curve averaged over 2500 grid points is also shown. As Fig.~7(b) demonstrates, the OPP tunneling current curves strongly depend on the nanoscale surface structures---this spatial dependence is inaccessible using the conventional OPP methods without STM. 

\begin{figure*}
\includegraphics[width=15cm]{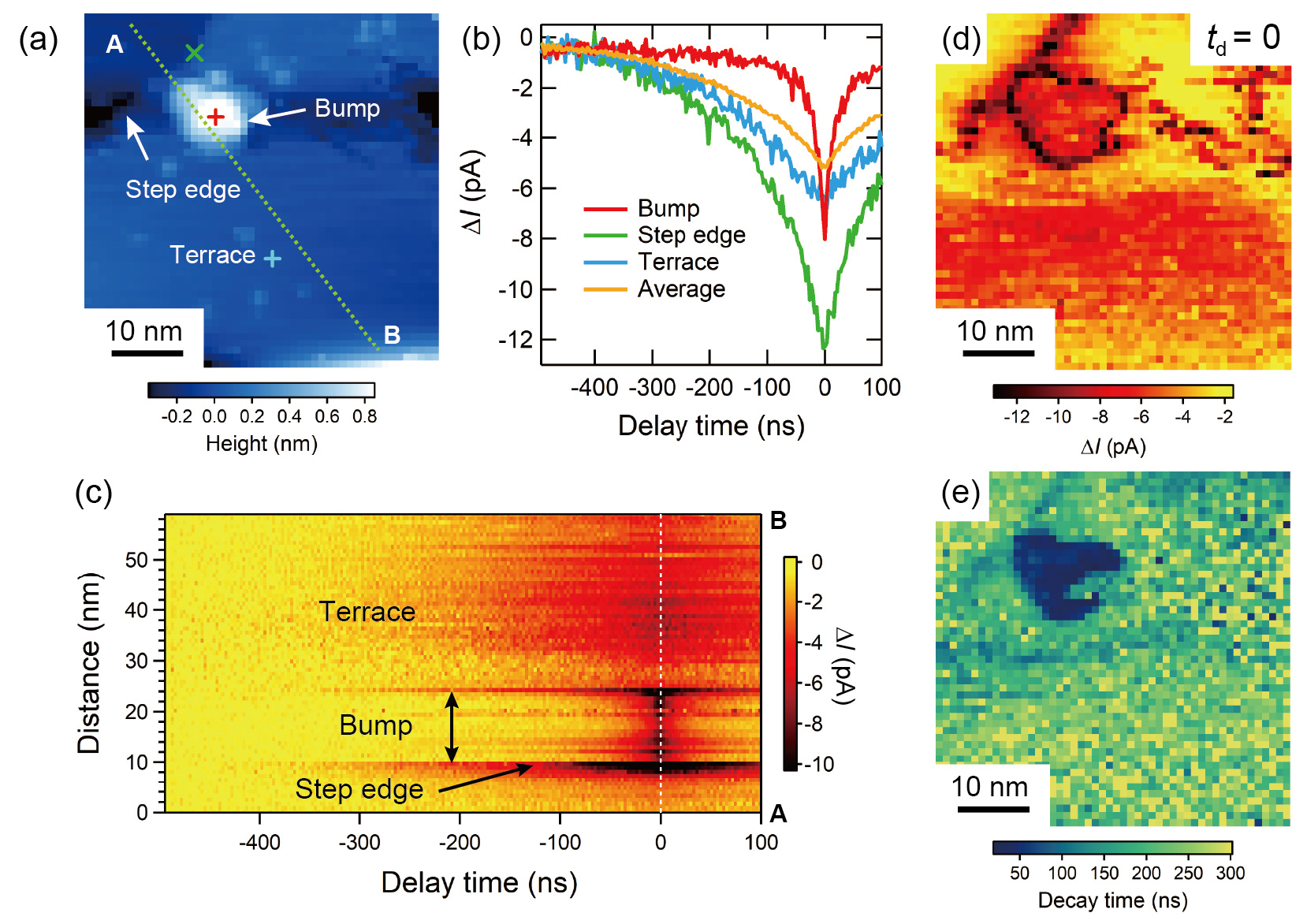}% Here is how to import EPS art
\caption{(a) STM image of GaAs(110) surface. Set-point: $V = +3$~V, $I_{\rm t} = 100$~pA. $T = 6$~K. (b) Characteristic spatially-dependent OPP tunneling current--delay time curves. The locations where the curves are taken are shown in (a) indicated with the same color. Laser power is 0.25~mW. (c) Line curves along A-B in (a). The step of delay time is 3~ns, and the averaging time at each delay time is 0.15~s in (b) and (c). (d) OPP tunneling current map at a delay time of 0~ns measured in the same field of view with (a). (e) Decay time map in the same field of view with (a). The decay time at each location is estimated by fitting the spectrum with a single exponential curve.
}
\label{fig7}
\end{figure*}

To examine the spatial dependence in more detail, we compile a series of line curves across the step edge, the bump structure, and the terrace. As Fig.~7(c) demonstrates, the OPP tunneling current curves not only identify the interior region of the bump structure, but they also identify the boundary between the bump and the terrace. In fact, the OPP tunneling current image at $t_{\rm d} = 0$ in Fig.~7(d) exhibits large amplitude $\Delta I$ both along the perimeter of the bump structure and along the step edge.

Furthermore, it is possible to map a decay time by fitting each curve with an exponential function. In this experiment, a relatively low laser power (0.25~mW) is chosen so that most of curves are well fitted with a single exponential function. The decay time map in Fig.~7(e) demonstrates that, inside the bump structure, the decay time ($\sim30-60$~ns) is substantially shorter than that of its surrounding. This short decay time might be attributed to the recombination of holes in the depletion layer with electrons injected from the tip into impurity states formed inside the bump structure, in analogy with the Co nanoparticle on GaAs(110) surface\cite{Terada_NatPhoton}. 

\begin{figure*}
\includegraphics[width=15cm]{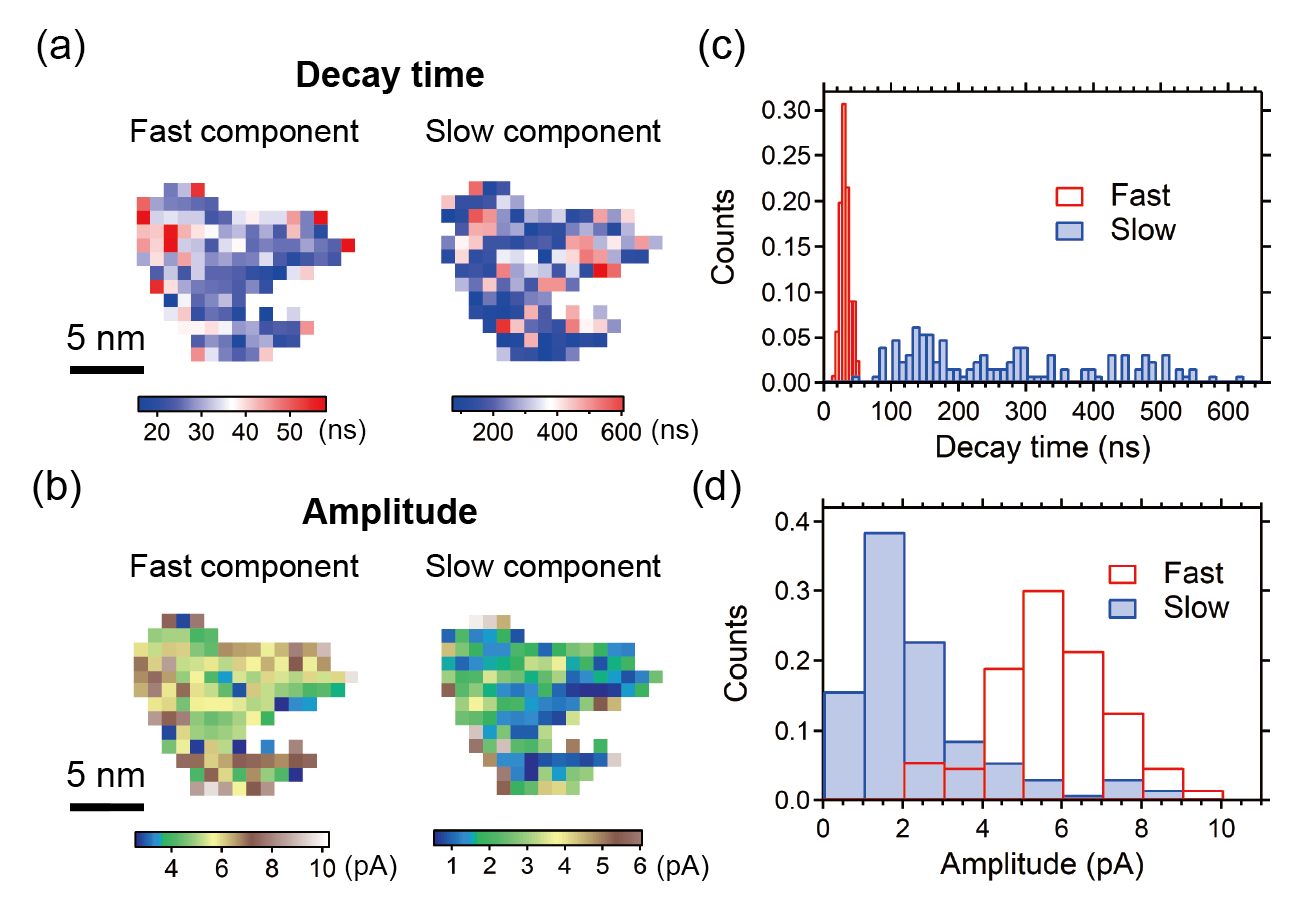}% Here is how to import EPS art
\caption{(a) Decay time map of the fast and slow components in the bump structure in Fig.~7(a). The decay time at each location is estimated by fitting the spectrum with double exponential curves. (Left) Fast component. (Right) Slow component. (b) Amplitude map of the double exponential curves. (Left) Fast component. (Right) Slow component. (c) Histogram of the decay time of the fast and slow components in (a). (d) Histogram of the amplitude of the fast and slow components in (b). The numbers of counts are normalized in (c) and (d).
}
\label{fig8}
\end{figure*}

Additionally, we fit OPP tunneling current curves taken inside the bump structure with two exponential functions, separating the fast and the slow components. Figures 8(a) and (b) show the maps of the decay time and of the amplitude of exponential function for each component, informing on the inhomogeneous spatial distribution of the fast and slow components. 
The histogram of the decay time shows the fast component ranges between 10 and 50~ns in a narrow temporal range [red in Fig.~8(c)], and the slow component ranges between 100 and 550~ns in a wider temporal range [blue in Fig.~8(c)]. The histogram of the amplitudes of the two components shows the fast component exhibits larger amplitudes than the slow component [Fig.~8(d)], possibly reflecting its dominant carrier dynamics. 
We note that the elucidation of the origin of these results requires samples with better defined nanoscale structures, which is beyond the scope of this work. Here, we emphasize the capability of the mapping techniques presented above for investigating ultrafast carrier dynamics with nanometer-scale spatial resolution and applicable for various semiconducting materials.

\begin{figure*}
\includegraphics[width=12cm]{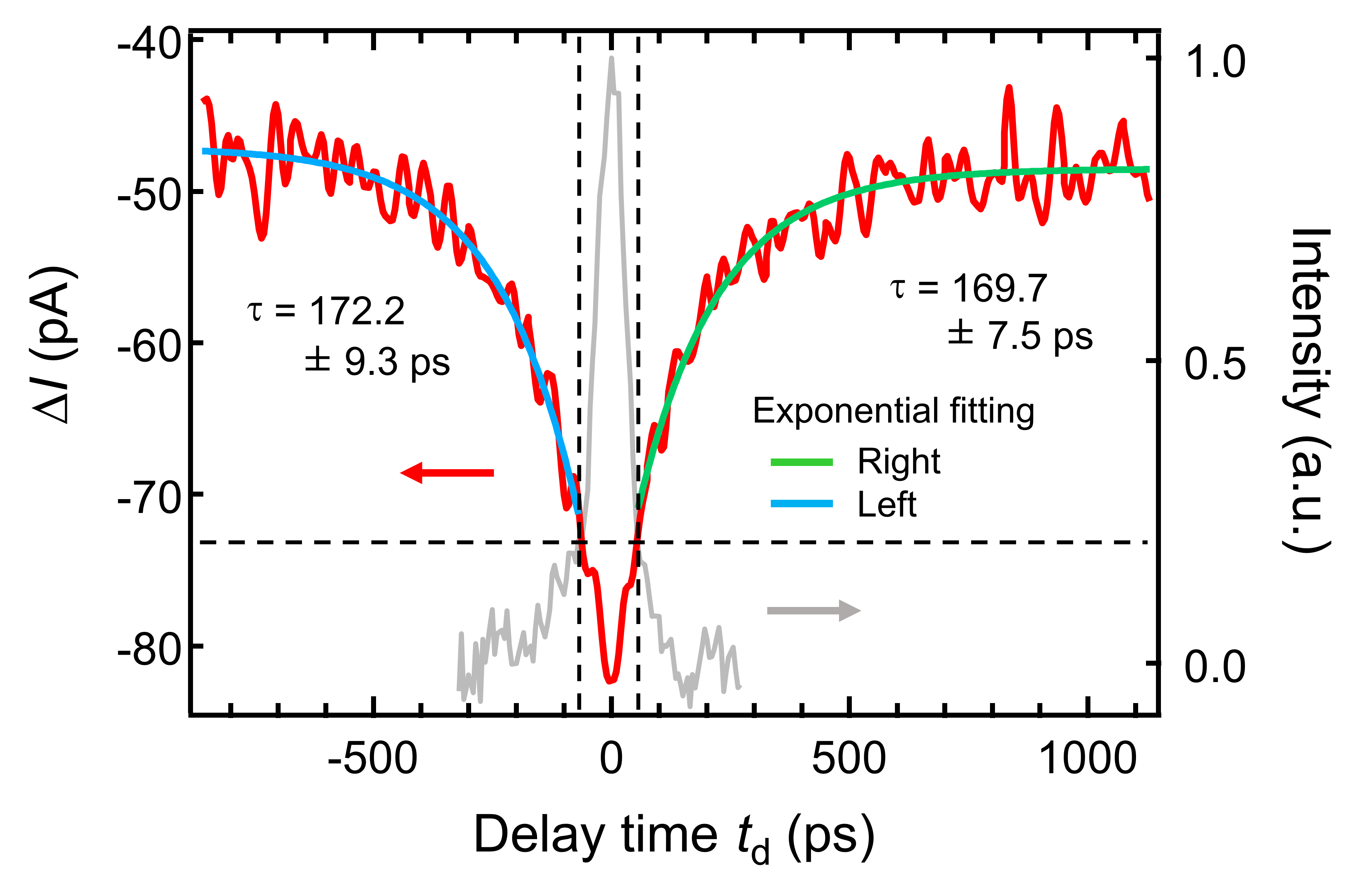}% Here is how to import EPS art
\caption{OPP tunneling current--delay time curve of LT-GaAs taken at room temperature. Set-point: $V = +5.5$~V, $I_{\rm t} = 1$~nA. Laser power is 4~mW. The step of delay time is 5~ps, and the averaging time at each delay time is 78~ms. The curve is averaged over 25 sweeps. The fast decay at around $t_{\rm d} = 0$ is attributed to the cross-correlation function between the pump and probe pulses (FWHM of $\sim 78$~ps) shown in Fig.~S2 (plotted on the right axis). The curves in the range of $|t_{\rm d}| > 55$~ps are fitted with an exponential function.
}
\label{fig9}
\end{figure*}

Finally, we conduct an OPP-STM measurement on  low-temperature-grown GaAs (LT-GaAs) to 
directly demonstrate the tens-picosecond range temporal resolution (Fig.~9). The fast decay at around $t_{\rm d} = 0$ is attributed to the cross-correlation between the pump and probe laser pulses (FWHM of $\sim 78$~ps) as shown in Supplementary Fig.~S2 online. By fitting the OPP tunneling current curve in the range of $|t_{\rm d}| > 55$ ps, we obtain the decay time of $\sim170$~ps, which could be originated from surface defect levels. The detail is beyond the scope of this work and will not be discussed here. This result clearly shows that the OPP-STM system developed in this study enables the detection of carrier dynamics in the tens-picosecond range significantly faster than the previous externally-triggerable OPP-STM systems\cite{Guo, Mogi_FPGA, Kloth_RSI}.

\section{Discussion}
We discuss the future possibilities of the OPP-STM technique presented in this study. 
First, the temporal resolution in the tens-picosecond range will greatly expand the range of applications. For example, the picosecond range exciton dynamics in transition metal dichalcogenides\cite{Mogi_npj} can be investigated using this system. Second, the optical system has significantly improved the stability of the laser illumination of the tunnel junction. The long-term stability of the laser illumination allows us to perform OPP tunneling current mapping, thereby visualizing carrier dynamics in the form of, for example, decay time maps as shown in Figs.~7 and 8. This mapping technique can be applied to reveal carrier dynamics associated with nanoscale structures such as domain boundaries in transition metal dichalcogenides and organic solar cell materials\cite{Takeuchi_APEX}. 
Photo catalysis\cite{Guo} and photo induced phase transition\cite{Terada_NanoLett} are also interesting phenomena to be investigated. Since a wide delay time range is available from ps to $\mu$s by adjusting the timing of electric pulses, the current system enables the capture of various photo-induced phenomena in these systems.  

Because in this work, the laser wavelength is fixed at 532~nm (2.33~eV), the selection of a laser system according to the band energy gap of the sample of interest is still necessary. The externally-triggerable picosecond laser systems with a wide range of wavelengths (532-1550~nm) are currently available\cite{Laser}. Cutting-edge laser technology, which has been rapidly developing, may realize a higher-performance optical system enabling a wavelength-variable, externally-triggerable laser system with a shorter pulse width in the future.  

The spatial resolution can be further improved down to the atomic level, as already demonstrated by several groups\cite{Yoshida_APEX, Guo, Kloth_SciAdv}.
We note that a time-resolved STM system (temporal resolution of faster than 30~fs) consisting of almost the same configuration as Fig.~3(a) except for the lens has recently realized atomic-scale imaging under laser illumination\cite{Arashida_ACSPhoton}, indicating the potential capability of the atomic resolution with optical excitation in the OPP-STM system developed in this study.

By using circular polarized illumination, ultrafast spin dynamics can be probed with a spatial resolution of $\sim 1$~nm, as demonstrated in GaAs/AlGaAs quantum wells\cite{Yoshida_NatNanotech}. Although we still need to either tune the laser excitation energy to a spin splitting energy, or find a suitable material possessing the spin splitting energy matching with the laser excitation energy, it is possible to conduct such spin dynamics measurements based on the circular polarization modulation technique\cite{Yoshida_NatNanotech} by both adding a Pockels cell and a quarter-wave plate after each pump and probe laser, and by modifying the electronic circuit in the delay-time control system in Fig.~1. 
For example, previous spin dynamics measurement of Mn on GaAs(110) surface has reported a change in the spin lifetime with increasing Mn density, more surface-sensitive than the conventional OPP method, but spatially averaged the spin-related tunneling current over the surface\cite{Wang_PCCP}. The mapping technique presented in this study may resolve the spatial distribution of the spin lifetime of Mn atoms.   

Furthermore, applying the OPP-STM technique to multiprobe systems offers us great opportunities to study carrier dynamics of various nanoscale structures. For example, small island structures on an insulating substrate, which are inaccessible to a single probe STM, can be observed by using one tip as an electrode, while conducting STM measurements using the other tip. Such OPP-multiprobe techniques have already been developed and applied for investigating a monolayer island of a WSe$_2$/MoSe$_2$ heterostructure\cite{Mogi_multiprobe},  monolayer and bilayer WSe$_2$ islands grown on SiO$_2$ substrates\cite{Mogi_JJAP}, and exciton dynamics in an in-plane WS$_2$/WSe$_2$ heterostructure\cite{Mogi_npj}. 
The placement of a position-movable aspheric lens directly on the multiprobe stage will similarly improve the stability of laser spot on the sample surface, allowing us to conduct time-resolved mapping experiments. 

The optical system developed in this study is also applicable for time-resolved atomic force microscopy\cite{Schumacher}, in which the thermal expansion effect needs to be similarly suppressed. 
Since the long-term stability of the laser spot on the sample surface is commonly indispensable for other time-resolved techniques, the optical system reported in this study can be  applied for various time-resolved measurements to improve data quality.

\section{Conclusion}
This work reports the development of an externally-triggerable OPP-STM system capable of conducting tens-picosecond range, long-term time-resolved measurements. 
We achieve this both by controlling the laser pulse sequence electrically, and by placing the aspheric lens on the same stage with the STM head together with the use of the beam stabilization system. 
The optical system enables us to conduct OPP-STM measurements with a temporal resolution
of $\sim 80$~ps.
We successfully demonstrate the decay time of $\sim 170$~ps using LT-GaAs. 
The temporal resolution will be further improved by using laser systems with a shorter pulse width (currently 45~ps) and by deliberately selecting the surrounding electronics in the future. A wide range of the laser wavelength is available by using a suitable externally-triggerable laser system. Our OPP-STM mapping data reveals nanoscale carrier dynamics on GaAs(110) surface, offering the potential capabilities of this technique to a deeper understanding of carrier dynamics in various advanced functional materials.

\begin{acknowledgments}
The authors are grateful to Masaharu Sakai and Kazuhiko Kurita (UNISOKU) for their technical assistance. This work was supported by Adaptable and Seamless Technology Transfer Program through Target-driven R\&D (A-STEP), Japan Science and Technology Agency (JST), and a Grant-in-Aid for Scientific Research (17H06088, 20H00341, 22H00289) from Japan Society for the Promotion of Science.
\end{acknowledgments}

\section*{Author contributions statement}
K. I. and M. Y. carried out the experiments and the data analyses. H. H., H. M., S. Y., and O. T. designed the delay-time control system. Y. M. and H. S. supervised the project. K. I. wrote the manuscript with input from all authors.  

\section*{Availability of Data and Materials}
The datasets used and/or analysed during the current study available from the corresponding author on reasonable request.

\section*{Competing interests} 
The authors declare no competing interests.

\bibliography{apssamp}% Produces the bibliography via BibTeX.

\end{document}